\documentclass[aps,prc,twocolumn,superscriptaddress,showkeys,nofootinbib]{revtex4-1}
\makeatletter
\renewcommand\subsection{\@startsection{subsection}{2}{\z@}
  {-2.0ex\@plus -.6ex \@minus -.2ex}
  {0.8ex \@plus .2ex}
  {\normalfont\normalsize\itshape}}
\makeatother
\usepackage{amsmath,amssymb}

\begin{document}

\title{Finite-Size Scaling of Net-Proton Cumulants in Heavy-Ion Collisions: \\
Remarks on the Interpretation of a Recent Finite-Size Scaling Analysis}

\author{Roy A. Lacey}
\affiliation{Department of Chemistry, Stony Brook University, Stony Brook, NY 11794, USA}

\begin{abstract}

Finite-size scaling (FSS) provides a framework for investigating the possible presence of a critical end point (CEP) in the QCD phase diagram using fluctuation observables measured in relativistic heavy-ion collisions. A recent analysis reported a finite-size scaling representation of a susceptibility constructed from net-proton cumulants and interpreted the resulting scaling behavior as evidence for a CEP near $\mu_B \approx 625$ MeV \cite{FSSpaper}. 
This note examines several aspects of that scaling construction. These include the identification of the pseudorapidity acceptance window with the physical system size used in the finite-size scaling relations, the influence of acceptance-driven multiplicity scaling in the susceptibility used for the scaling analysis, and the treatment of thermodynamic scaling fields in the scaling variable. 
These considerations clarify several issues relevant for the consistent implementation and interpretation of finite-size scaling analyses in heavy-ion collision experiments. 
Taken together, they indicate that the reported scaling behavior does not uniquely establish critical dynamics nor the presence of a CEP near $\mu_B \approx 625$ MeV.

\end{abstract}

\maketitle

\subsection{Introduction}

The search for the critical end point (CEP) in the QCD phase diagram remains a central objective of relativistic heavy-ion collision experiments. 
Near the CEP, thermodynamic susceptibilities are expected to exhibit universal scaling behavior associated with the growth of the correlation length $\xi$ \cite{Stephanov:1998dy,Stephanov:1999zu}. Higher-order cumulants of conserved charges are particularly sensitive to this behavior and have therefore been proposed as experimental signatures of critical phenomena \cite{Stephanov:2008qz}.

Finite-size scaling (FSS) provides a well-established framework for identifying critical behavior in systems of limited spatial extent \cite{Fisher:1972zz,Barber:1983zz}. 
Because the fireball created in heavy-ion collisions has a finite size and lifetime \cite{Berdnikov:1999ph}, scaling relations may be tested experimentally using observables measured across different beam energies, collision systems, and centralities. Applications of finite-size scaling methods to fluctuation-derived observables in heavy-ion collisions have been explored in several studies \cite{Lacey:2014wqa,Lacey:2024mnv}.

Near a critical point the finite-size scaling form of a susceptibility $\chi$ is given by
\begin{equation}
\chi(L,r,h)=L^{\gamma/\nu}\Phi(rL^{1/\nu},hL^{\beta\delta/\nu}),
\label{eq:FSS}
\end{equation}
where $L$ denotes the characteristic spatial size of the system, $r$ and $h$ represent the temperature-like and ordering-field scaling variables, respectively, $\Phi$ is a universal scaling function, and $\nu$, $\beta$, $\gamma$, and $\delta$ are critical exponents of the underlying universality class. 
In this representation the dependence on the two independent scaling fields enters through the combinations $rL^{1/\nu}$ and $hL^{\beta\delta/\nu}$, reflecting the distinct roles of the temperature-like and ordering-field directions in the vicinity of a critical point. 
When the appropriate scaling variables are used and the physical system size is varied, Eq.~(\ref{eq:FSS}) predicts that suitably rescaled observables measured for different system sizes collapse onto a universal scaling function.

In heavy-ion collisions the thermodynamic coordinates associated with the scaling fields $r$ and $h$ are commonly expressed in terms of combinations of temperature $T$ and baryon chemical potential $\mu_B$. The physical system size entering the finite-size scaling relations is instead determined by the spatial extent of the fireball produced in the collision, which depends primarily on the nuclear geometry and the collision centrality.
The interpretation of such scaling patterns therefore requires careful consideration, since the physical system size, detector acceptance, and thermodynamic parameters entering the scaling variables must be clearly distinguished when applying finite-size scaling methods to experimental observables.

A recent analysis reported a FSS collapse of a susceptibility constructed from net-proton cumulants that was interpreted as evidence for a CEP near $\mu_B \approx 625$ MeV \cite{FSSpaper}. In that work the scaling construction combines the dependence of measured cumulants on the pseudorapidity acceptance window with thermodynamic parameters obtained from chemical freeze-out analyses, thereby introducing several assumptions regarding the physical system size, acceptance dependence, and thermodynamic scaling variables.

This note examines several assumptions underlying this scaling construction. 
These include the identification of the pseudorapidity acceptance window with the physical system size used in the finite-size scaling relations, the role of acceptance-driven multiplicity scaling in the constructed susceptibility, and the treatment of thermodynamic scaling fields in the scaling variable.

\subsection{System size in finite-size scaling}

Finite-size scaling relates thermodynamic observables to the spatial size of a system near a critical point. 
In statistical mechanics the relevant size parameter $L$ represents the linear extent of the system that limits the growth of the correlation length $\xi$.

In relativistic heavy-ion collisions the physical system size is determined primarily by the geometric overlap region of the colliding nuclei and the subsequent evolution of the produced fireball. 
The characteristic transverse scale may be approximated as $L \sim R_\perp$, where $R_\perp$ denotes the transverse radius of the participant zone. 
In Glauber descriptions this scale increases approximately with the number of participating nucleons as $R_\perp \propto N_{\rm part}^{1/3}$, so that variations in centrality or comparisons among collision systems naturally provide a means of varying the physical system size relevant for finite-size scaling.

It is worth noting that the finite-size scaling illustration presented in Fig.~1 of Ref.~\cite{FSSpaper} employs subsystems of different linear size $L$ in a model calculation where the control parameter is the baryon density $n_b$. 
In that construction $L$ represents the true spatial extent of the subsystem and therefore limits the growth of the correlation length. 
The scaling variable is therefore density-driven, corresponding to a temperature-like scaling direction, with the scaling combination entering through $(n_b-n_c)L^{1/\nu}$, where $n_c$ denotes the critical baryon density. 
This form is consistent with the $rL^{1/\nu}$ scaling variable appearing in Eq.~(\ref{eq:FSS}).

In the analysis of the experimental data in Ref.~\cite{FSSpaper}, the pseudorapidity acceptance window $W$ is instead treated as the system size entering the finite-size scaling relations. 
However, varying the pseudorapidity acceptance modifies only the fraction of particles included in the measurement rather than the spatial extent of the fireball produced in the collision. 
For a fixed collision system and centrality selection the geometric size of the system—determined by the nuclear overlap region—and therefore the spatial bound on the correlation length remain unchanged.

Consequently, changes in the acceptance window correspond to variations of the measurement volume rather than the thermodynamic system size that controls the growth of the correlation length. 
Finite-size scaling arises because the spatial system size limits the growth of the correlation length, $\xi \le L$. 
Changing the rapidity acceptance window therefore does not alter the physical system size relevant for finite-size scaling.

Such variations are nevertheless known to influence measured cumulants through multiplicity scaling, binomial acceptance sampling, and global conservation effects \cite{BraunMunzinger:2011ta,Bzdak:2012aa,Bzdak:2012ab,Bzdak:2012an,Kitazawa:2012at}. 
While these mechanisms introduce a smooth dependence of cumulants on the pseudorapidity interval, they do not modify the spatial constraint that governs the finite-size scaling behavior.

\subsection{Acceptance-driven multiplicity scaling}

Even if the rapidity acceptance window is provisionally treated as a scaling variable, the construction of the observable used in Ref.~\cite{FSSpaper} can itself promote a collapse of the plotted observables. 
Because cumulants scale with the number of particles entering the measurement, their dependence on the rapidity window can generate systematic scaling behavior even in the absence of critical dynamics. 
This acceptance dependence can therefore imprint a systematic scaling pattern on observables constructed from the measured cumulants.

For approximately uncorrelated particle production the variance of the net-particle distribution scales with the number of particles entering the measurement, $C_2 \propto N$. 
Since the multiplicity increases approximately linearly with the pseudorapidity interval, $N \propto W$, the leading acceptance dependence of the second cumulant is therefore $C_2 \propto W$. 
Similar acceptance-driven behavior can also arise from global conservation effects and binomial acceptance sampling \cite{Bzdak:2012aa,Bzdak:2012ab,Kitazawa:2012at}.

In Ref.~\cite{FSSpaper} the susceptibility used in the scaling analysis is defined as
\begin{equation}
\chi_2=\frac{C_2}{T_{fo}^3\,W\,dV_{fo}/dy}.
\end{equation}
Substituting the approximate multiplicity scaling $C_2 \propto W$ into Eq.~(2) shows that the dominant dependence on the rapidity window cancels, yielding $\chi_2 \sim W/W \approx \mathrm{const}$ to leading order. 
This cancellation occurs prior to the introduction of any critical exponents and therefore reflects the construction of the observable rather than critical scaling.

The observable used in the scaling representation therefore behaves approximately as
\begin{equation}
\chi_2 W^{-\gamma/\nu} \propto W^{-\gamma/\nu}.
\end{equation}
Because the leading acceptance dependence cancels in the constructed observable, the normalization used in the scaling representation naturally promotes a collapse of the plotted data when expressed in the proposed scaling variables. 
The resulting scaling behavior therefore follows directly from the acceptance dependence of the observable rather than from the presence of critical correlations.

\subsection{Scaling fields near the critical point}

Thermodynamic singularities near a critical point are governed by two independent scaling fields. 
For QCD these fields correspond to combinations of temperature and baryon chemical potential, commonly expressed as the temperature-like variable $r \sim (T-T_c)/T_c$ and the ordering-field variable $h \sim (\mu_B-\mu_{B,c})/\mu_{B,c}$, where $T_c$ and $\mu_{B,c}$ denote the temperature and baryon chemical potential at the critical end point. 
Finite-size scaling analyses therefore require consistent identification of both thermodynamic scaling fields and the critical exponents governing their scaling behavior in the combinations $rL^{1/\nu}$ and $hL^{\beta\delta/\nu}$.

In the analysis of Ref.~\cite{FSSpaper} the thermodynamic scaling variable is effectively reduced to the baryon chemical potential through the combination $(\mu_B-\mu_{B,c})/\mu_{B,c}$, while the second thermodynamic scaling field associated with temperature or baryon density is not explicitly incorporated. This variable is then associated with the finite-size scaling exponent $1/\nu$. However, in the mapping of QCD thermodynamics to the three-dimensional Ising universality class the baryon chemical potential is generally associated with the ordering-field direction. In finite-size scaling theory the exponent governing the scaling variable depends on the thermodynamic direction being probed: the temperature-like field enters through the combination $rL^{1/\nu}$, whereas the ordering field enters through $hL^{\beta\delta/\nu}$. 
Associating $(\mu_B-\mu_{B,c})/\mu_{B,c}$ with the exponent $1/\nu$ therefore mixes the temperature-like and ordering-field directions and further obscures the interpretation of the resulting scaling representation.

A further complication arises from the dependence of the constructed susceptibility on model-extracted freeze-out parameters. The susceptibility used in Ref.~\cite{FSSpaper} incorporates the freeze-out temperature $T_{fo}$ and the freeze-out volume density $dV_{fo}/dy$, which are obtained from statistical hadronization fits rather than direct measurements. 
Because these quantities depend on the specific thermodynamic model used to describe chemical freeze-out, the resulting susceptibility inherits an additional level of model dependence that is not present in the experimentally measured cumulants themselves.

\subsection{Choice of observable}

The search for critical behavior in heavy-ion collisions, including finite-size scaling studies, typically relies on fluctuation observables constructed from the cumulants of conserved-charge distributions and their ratios. 
The scaling analysis of Ref.~\cite{FSSpaper} focuses on a susceptibility derived from the second-order cumulant of the net-proton distribution. 
While this observable provides useful information on baryon-number fluctuations, searches for critical behavior generally benefit from examining several fluctuation observables simultaneously, including higher-order cumulants and their ratios.

Near a critical point the cumulants of conserved charges are expected to exhibit characteristic scaling with the correlation length $\xi$ \cite{Stephanov:2008qz}. 
Parametrically one finds $C_2 \sim \xi^2$, $C_3 \sim \xi^{4.5}$, and $C_4 \sim \xi^{7}$, indicating that higher-order cumulants probe progressively stronger powers of the correlation length. 
At the same time, the finite size and finite lifetime of the fireball produced in heavy-ion collisions limit the growth of the correlation length, so dynamical effects may influence higher moments more strongly. 
These differing powers of the correlation length also imply distinct sign and magnitude patterns for the cumulant ratios near the critical region.

Cumulant ratios provide an additional advantage because they reduce the leading dependence on the system volume and acceptance. 
The ratio $C_2/C_1$, often referred to as the scaled variance, is closely related to the baryon-number susceptibility and is expected to increase as the correlation length grows near a critical point. 
Higher-order ratios such as $C_3/C_2$ and $C_4/C_2$ probe progressively stronger powers of the correlation length and therefore provide enhanced sensitivity to critical dynamics. 
Importantly, these ratios exhibit distinct divergence patterns in the vicinity of the critical point. 
The skewness ratio $C_3/C_2$ is sensitive to the ordering-field direction and can change sign as the system passes through the critical region, while the kurtosis ratio $C_4/C_2$ can display pronounced non-monotonic behavior and become negative over portions of the critical domain. 
By contrast, ratios such as $C_4/C_1$ may diverge positively as the correlation length increases. 
These differing divergence patterns reflect the underlying universality constraints governing critical fluctuations and therefore provide important consistency tests for identifying the presence and location of a critical point. 
Recent finite-size scaling studies that incorporate cumulant ratios illustrate the constraining power of such combined observables and can yield CEP locations that differ from those inferred using a single susceptibility-based scaling analysis \cite{Lacey:2024mnv}.

Consequently, the interpretation of fluctuation measurements in terms of critical phenomena generally relies on consistent behavior observed across multiple cumulants and cumulant ratios rather than on a single observable. 
The combined growth patterns, sign structure, and scaling behavior of these observables provide important consistency checks that help distinguish potential critical contributions from acceptance effects, volume fluctuations, and other noncritical sources of correlations.

\subsection{Summary}

The finite-size scaling analysis reported in Ref.~\cite{FSSpaper} represents an interesting attempt to apply critical scaling concepts to fluctuation measurements in relativistic heavy-ion collisions. 
However, several methodological aspects of the analysis complicate the interpretation of the reported scaling behavior as evidence for a CEP near $\mu_B \approx 625$ MeV.

First, the pseudorapidity acceptance window is identified with the system size entering the scaling relations even though variations of the acceptance window modify only the detector acceptance rather than the spatial extent of the fireball. 
In heavy-ion collisions the physical system size is instead determined primarily by the geometric overlap region of the colliding nuclei and varies with centrality and collision system.

Second, the susceptibility used in the scaling representation combines measured cumulants with model-extracted freeze-out parameters and the rapidity acceptance window. 
Because cumulants scale with the number of particles entering the measurement, this construction can promote an apparent scaling pattern in the plotted observables through acceptance-driven multiplicity scaling.

Third, the scaling variable effectively reduces the thermodynamic dependence to a single dimension involving the baryon chemical potential, while the temperature-like scaling field required in critical scaling theory is not explicitly incorporated. 
This also leads to an association of the scaling variable with the exponent $1/\nu$, even though the baryon chemical potential is generally related to the ordering-field direction in the mapping of QCD thermodynamics to the three-dimensional Ising universality class.

Finally, the interpretation of fluctuation measurements in terms of critical phenomena generally benefits from examining several cumulants and cumulant ratios simultaneously. 
Ratios such as $C_2/C_1$, $C_3/C_2$, $C_4/C_2$, and $C_4/C_1$ probe different powers of the correlation length and exhibit distinct divergence patterns near the critical region, providing complementary constraints on the presence and location of a critical point.

Taken together, these considerations indicate that the reported scaling behavior does not uniquely establish critical dynamics or the presence of a CEP near $\mu_B \approx 625$ MeV inferred in Ref.~\cite{FSSpaper}. 
Finite-size scaling remains a powerful framework for identifying critical phenomena when the physical system size and thermodynamic scaling fields are varied consistently with the underlying scaling theory and when consistent signatures are observed across multiple fluctuation observables.

\end{document}